\documentclass{jltp}

\usepackage{graphicx}
\usepackage{amssymb}

\def\be{\begin{eqnarray}}
\def\ben{\begin{eqnarray*}}
\def\ee{\end{eqnarray}}
\def\een{\end{eqnarray*}}

\def\eps{\varepsilon}
\def\n{{\bf n}}
\def\m{{\bf m}}
\def\x{{\bf x}}
\def\z{{\bf 0}}
\def\avec{\hat{\bf a}}
\def\alphavec{\hat{\mbox{\boldmath{$\alpha$}}}}
\def\calE{{\cal{E}}}
\def\JE{\mathfrak{E}}

\begin{document}

\title{Weakly-Interacting Bosons in a Trap within Approximate Second Quantization Approach}

\author{Andrij Rovenchak}

\address{Department for Theoretical Physics,
Ivan Franko National University of Lviv,\\
12 Drahomanov St., Lviv, UA--79005, Ukraine}

\runninghead{A. Rovenchak}{Trapped weakly-interacting bosons in approximate second quantization}

\maketitle

\begin{abstract}
The theory of Bogoliubov is generalized for the case of a weakly-interacting Bose-gas in
harmonic trap. A set of nonlinear matrix equations is obtained to make the
diagonalization of Hamiltonian possible. Its perturbative solution is used for the
calculation of the energy and the condensate fraction of the model system
to show the applicability of the method.

\end{abstract}

\section{INTRODUCTION}\label{sec1}

In recent decades,
the experimental achievements in the investigation of
ultra-cold alkali gases\cite{And95,Dav95} revived the interest to theoretical models
of bosonic systems\cite{Pet01}.
The quantum field-theoretical techniques occupy a predominant position within the
methods for the studies of weakly-interacting bosons.
The creation--annihilation operator formalism was directly applied in Refs.~\onlinecite{Bog47,Bas64,Oku03}.
The use of nonlinear Schr\"odinger equation (Gross--Pitaevskii equation) is much more popular
for the studies in this domain\cite{Dal99}.
Some modifications\cite{HFu03} and extensions\cite{Kol00} of this methods are also available.
Numerical techniques were utilized by Krauth\cite{Kra96}, Pearson et al.\cite{Pea98} and may others.
The low-dimensional Bose-systems became of special interest recently\cite{Pea98,Dru97,Khe05,Bat05}.

In this work, a method similar to that of Bogoliubov\cite{Bog47} is
developed for
a weakly-interacting $D$-dimensional Bose-gas in the harmonic trapping potential.
The paper is organized as follows. In the next Section,
the Hamiltonian of the system under study is written in the approximate
second quantization approach. As a result of its diagonalization a set of matrix equations is obtained.
The results of calculations of the condensate fraction and energy are given in Section \ref{sec6}

\section{HAMILTONIAN DIAGONALIZATION}\label{sec2}
We consider a $D$-dimensional system of $N$ weakly-interacting bosons of mass $m$
confined in the harmonic trap
\be
V(x_1,\ldots,x_D)={m\over2}\left(\omega_1^2 x_1^2+\ldots+\omega_D^2
x_D^2\right).
\ee
The potential of interatomic
interaction is given by
$U(x_1,\ldots,x_D)=g\delta(\x)$,
where the vector $\x=(x_1,\ldots,x_D)$, $g$ is the interaction
strength.

The Hamiltonian reads
\be \label{H1}
\hat H = \sum_{j=1}^N \left[{\hat{\bf p}_j^2\over 2m}+V(\x_j)\right]
+ \sum_{1\leq j<l\leq N} U(\x_j-\x_l) = \hat H_0+\hat U.
\ee
Here, $\hat{\bf p}_j$ is the momentum operator of the $j$th particle, $\x_j$ is its coordinate.

It is possible to develop the second quantization approach using the eigenfunctions
$|\n\rangle=|n_1,\ldots,n_D\rangle$ of
the operator $\hat H_0$, i.\,e., an ordinary $D$-dimensional harmonic oscillator.
For simplicity, we further demand that no ratio of the frequencies
$\omega_1,\ldots,\omega_D$ is a rational number to avoid an accidental
degeneration of energy levels.

Let $\hat a^\dag_\n, \hat a_\n$ be the creation and annihilation operators for the
state $|\n\rangle$.
The corresponding energy levels are
$
\eps_\n = \hbar\left(\omega_1n_1+\ldots+\omega_Dn_D\right).
$
In this representation the Hamiltonian (\ref{H1}) is
\be \label{H2}
\hat H = \sum_{\n}\eps_n \hat a^\dag_\n \hat a_\n
{}+{1\over2}\sum_{\m,\m',\n,\n'}
\langle \m\n|U|\m'\n'\rangle \hat a^\dag_\m \hat a^\dag_\n \hat a_{\m'}\hat a_{\n'}.
\ee
The operators satisfy standard bosonic commutation relations:
\be \label{commut}
\left[\hat a_{\n'}, \hat a^\dag_\n\right]=\delta_{\n\n'}
\ee

Now we apply an approximate second quantization procedure following Bogoliubov\cite{Bog47}.
Let $N_0$ be the number of particles at the lowest energy level $\eps_0$.
As the interaction is weak, the behaviour of bosons does not differ much from
that of an ideal system. That is, one can expect the Bose--Einstein condensation to occur
at low temperatures. The number $N_0$ is thus a macroscopic number.
As it is an eigenvalue of the operator $\hat a^\dag_0\hat a_0$, one can treat
$\hat a^\dag_0$ and $\hat a_0$ as $c$-numbers:
$
\hat a^\dag_0\hat a_0=N_0,\  \hat a_0\hat a^\dag_0=N_0+1\simeq N_0,\quad
\hat a^\dag_0\simeq\sqrt{N_0},\  \hat a_0\simeq\sqrt{N_0}
$.
To obtain more general results we do not put here $N_0=N$.
Note, that for $D<3$ the condensation appears
only in traps\cite{Bag91}.

Further, following Bogoliubov, we neglect the terms having more
than two operators with non-zero index.
As the eigenfunctions $|\n\rangle$ are real
the matrix elements in the second item of (\ref{H2}) at different operator combinations
are equal:
$
\langle \m\z|U|\n\z\rangle =
\langle \m\n|U|\z\z\rangle =
\langle \z\z|U|\m\n\rangle \equiv g\,c_{\m\n}
$,
where
\be
 c_{\m\n} = \left(m\omega\over2\pi^2\hbar\right)^{D/2}\prod_{j=1}^D (-1)^{(3m_j+n_j)/2}
 {1\over\sqrt{m_j!\,n_j!}}\,\Gamma\left(m_j+n_j+1\over2\right)
\ee

\vspace*{-0.1cm}
\noindent
if $m_j+n_j$ is even for all $j$ and $c_{\m\n} = 0$
otherwise.
The notation $\omega=(\omega_1\!\ldots\omega_D)^{1\over D}$.

The matrix elements with three zeros equal
\be
\langle\n\z|U|\z\z\rangle \equiv g\,d_\n =
g\left(m\omega\over2\pi^2\hbar\right)^{D/2}\prod_{j=1}^D (-1)^{n_j/2}
 {1\over\sqrt{n_j!}}\,\Gamma\left(n_j+1\over2\right)
\ee

\vspace*{-0.1cm}
\noindent
if $n_j$ is even for all $j$.

The Hamiltonian (\ref{H2}) becomes
\be \label{H4}
\hat H &=& {\rm const}+\sum_\n\eps_\n \hat a^\dag_\n \hat a_\n+
gN_0^{3/2}\sum_\n d_\n \left(\hat a^\dag_\n + \hat a_{\n}\right)
\nonumber
\\
&&{}+g\,{N_0\over2}\sum_{\m,\n} c_{\m\n} \left( 4\hat a^\dag_\m
\hat a_{\n} +  \hat a^\dag_\m \hat a^\dag_{\n} + \hat a_\m \hat
a_{\n} \right),
\ee
where `const' denotes the items of a
non-operator nature. Hereafter, it will be dropped. For brevity,
the conditions $\n\neq0,\m\neq0$ is not written explicitly.
Note the appearance of linear terms $\sim \hat a^\dag, \sim \hat a$ which were absent
in Bogoliubov's approach\cite{Bog47} due to the momentum conservation law.

Let the indices $m,n$ run over all the states denoted by vector indices $\m,\n$.
In order to obtain an energy spectrum of the Hamiltonian from Eq.~(\ref{H4}) we will try to diagonalize
it.
For this purpose, it is possible to write (\ref{H4}) in the following matrix form:
\be \label{H6}
\hat H = \avec^\dag \calE \avec + 2\lambda\sqrt{N_0} \left(\avec^\dag{\bf d} + \avec^T{\bf d} \right) +
\lambda \left( 4\avec^\dag C \avec + \avec^\dag C \avec^\dag {}^T + \avec^T C \avec \right)
\ee
In the above equation, $\avec$ and ${\bf d}$ are vectors of infinite dimension:
$$
\avec^T = \left(\hat a_1,\, \hat a_2\, \ldots \right),
\qquad
\avec^\dag = \left(\hat a^\dag_1,\, \hat a^\dag_2,\;
\ldots\right),
\quad
{\bf d}^T = \left(d_1,\, d_2,\, \ldots\right),
$$
The diagonal matrix
$
\calE = \left(\begin{array}{cccc} \eps_1 & 0 & 0 &\ldots\\
                                     0   & \eps_2 & 0 & \ldots \\
                                  \vdots &        & \ddots \\
        \end{array}\right),
$
and the matrix elements of $C$ are the coefficients $c_{mn}$.
\ We have also written $g\,{N_0/2} = \lambda$ for brevity.

In order to obtain the Hamiltonian in a diagonal form,
\be \label{H5'}
\hat H = \sum_n \epsilon_n \hat \alpha^\dag_n \hat\alpha_n  = \alphavec^\dag \JE\, \alphavec,
\ee
where $\JE$ is the diagonal matrix with elements $\epsilon_n$, one can apply a generalization of the well-known
Bogoliubov's $u$--$v$ transformation demanding
\be
\avec = X \alphavec + Y \alphavec^\dag {}^T + {\bf z}, \qquad
\avec^\dag = \alphavec^\dag X + \alphavec^T Y + {\bf z}^T.
\ee

\vspace*{-0.1cm}
\noindent
Here, $X$ and $Y$ are square matrices of infinite dimension, we will require them to be Hermitian (symmetric and
real), and ${\bf z}$ is a vector with real components.
From the commutation relation (\ref{commut}) one can show that matrices $X$ and $Y$
obey the following condition
\be \label{matrixcommut}
X^2 - Y^2 = I,
\ee

\vspace*{-0.3cm}
\noindent
where $I$ is a unit matrix.

It is easy to show that the linear terms (the expressions in the first parenthesis)
of (\ref{H6}) produce only a shift of the energy levels but do not
affect the distance between them. That is why hereafter we consider only the quadratic in $\avec$ and $\alphavec$
terms of the Hamiltonian.

To eliminate the linear in $\alphavec$ terms from $\hat H$, the
vector $\bf z$ must be
\be
{\bf z} = -2\lambda\sqrt{N_0}\,(\calE + 6\lambda C)^{-1} {\bf d}.
\ee

Demanding the terms $\hat\alpha^\dag_m\hat\alpha^\dag_n$ and $\hat\alpha_m\hat\alpha_n$ to vanish in the Hamiltonian
one gets the following matrix equations:
\be
&&X\calE Y + 4\lambda XCY + \lambda XCX + \lambda YCY = 0, \label{matrixeqs1}\\
&&Y\calE X + 4\lambda YCX + \lambda XCX + \lambda YCY = 0. \label{matrixeqs2}
\ee

The matrix $\JE$ is
\be
\JE = X\calE X + Y\calE Y + 4\lambda(XCX+YCY) + 2\lambda(XCY + YCX)
\ee

\vspace*{-0.1cm}
\noindent
and its eigenvalues define the energy spectrum.

If the interatomic interaction is turned off ($g=0$), the solutions are
$
X=I,\  Y=0.
$
We will expand the matrices $X$ and $Y$ into series over $\lambda$:
\be
X=I+2\lambda^2\chi^2 +\ldots,\qquad Y=\lambda\upsilon + \lambda^2\upsilon_1
+\ldots\,\,.
\ee
The matrices $\chi,\upsilon,\upsilon_1$ can be found from
Eqs.~(\ref{matrixcommut})--(\ref{matrixeqs2}):
\be
\upsilon = 2\chi;\qquad \chi=-\calE^{-1}C/2;\qquad
\upsilon_1 = 4\,\calE^{-1}C\calE^{-1}C.
\ee

In the approximation up to $\lambda^2$ we obtain the matrix $\JE$:
\be \label{perturbE}
\JE = \calE + 4\lambda C + {\lambda^2\over2}\,
\left\{ \left(\calE^{-1}C\right)^2\calE - 3\,C\calE^{-1}C - 2 \calE^{-1}C^2 \right\}
\ee
and the energy levels 
are given by
$
\epsilon_n = \eps_n + 4\lambda c_{nn} + {\cal O}(\lambda^2).
$

Equations (\ref{matrixcommut})--(\ref{matrixeqs2}) are
difficult to solve, being non-linear with respect to infinite matrices.
This set may be treated as coupled algebraic (matrix) Riccati equations:

\vspace*{-0.8cm}
\be
&&XAY + XBX + YBY = 0,\\
&&YAX + XBX + YBY = 0,\\
&& X^2 - Y^2 - I = 0,
\ee

\vspace*{-0.2cm}
\noindent
where
$
A = \calE + 4\lambda C,\ B = \lambda C.
$
The numerical (non-perturbative) solution of this set needs special approaches\cite{RicattiEqs}.
This problem will be considered in a separate paper.

\vspace*{-0.35cm}
\section{CALCULATION RESULTS}\label{sec6}

If Bose--Einstein condensation occurs, the chemical potential $\mu$
of the system approaches zero and the total number of particles is
given by
\be
N = N_0 + \sum_{n>0} {1\over \exp(\epsilon_n/T) - 1},
\ee
%
where $N_0$ is the occupation of the lowest energy level.
Therefore, given $N$ and the energy spectrum
$
\epsilon_n = \epsilon_n(\lambda) = \epsilon_n(gN_0)
$
from the eigenvalues of (\ref{perturbE}),
one can calculate $N_0$ as a function of temperature $T$. We thus arrive at a self-consistent
problem.
Its solution allows one to calculate the energy
\be
E = E_0 + \sum_{n>0} {\epsilon_n\over \exp(\epsilon_n/T) - 1}
\ee

\vspace*{-0.1cm}
\noindent
and other thermodynamic functions.

\begin{figure}[h]
\centerline{\includegraphics[scale=.25,clip]{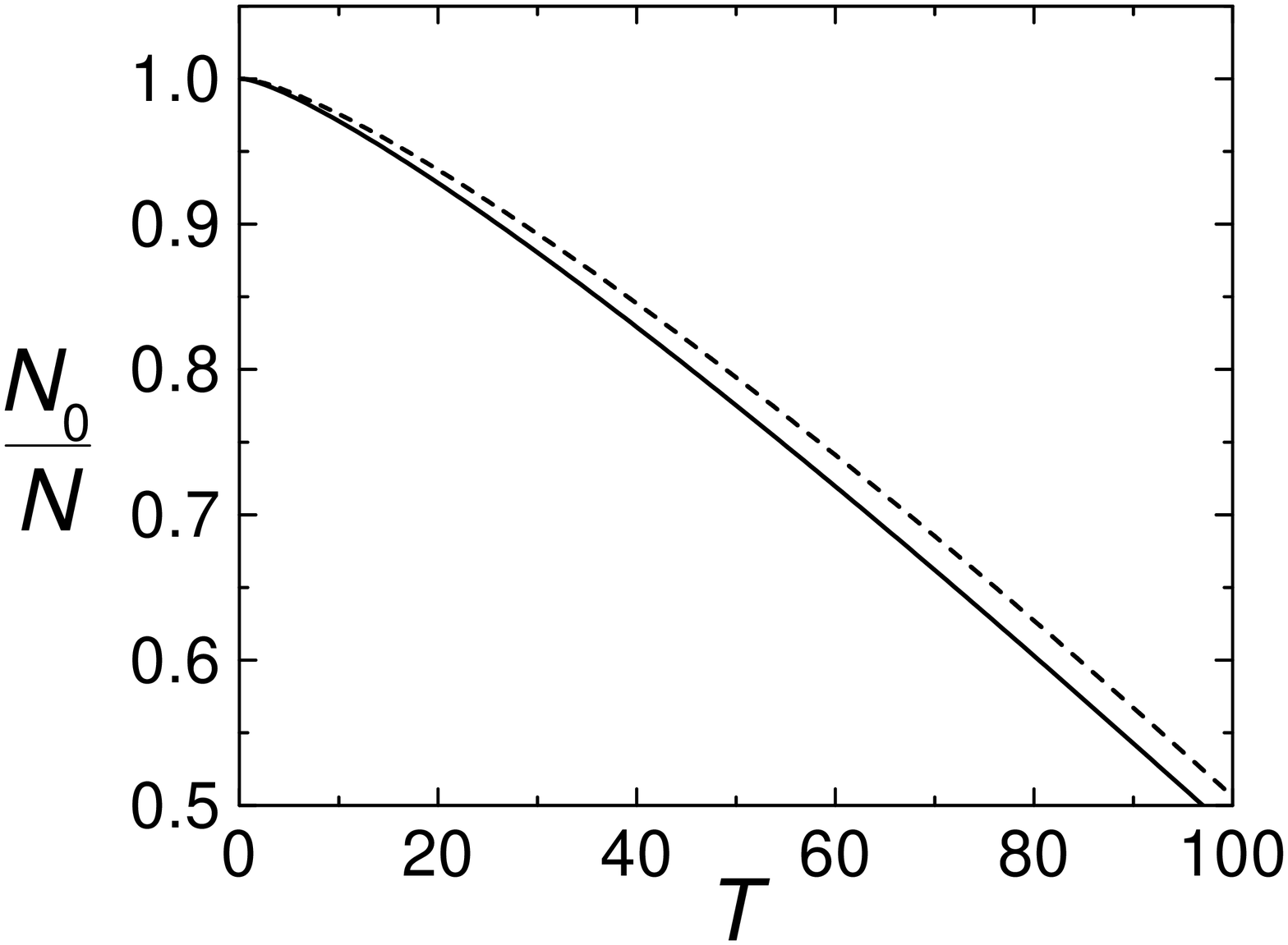}\qquad\includegraphics[scale=.25,clip]{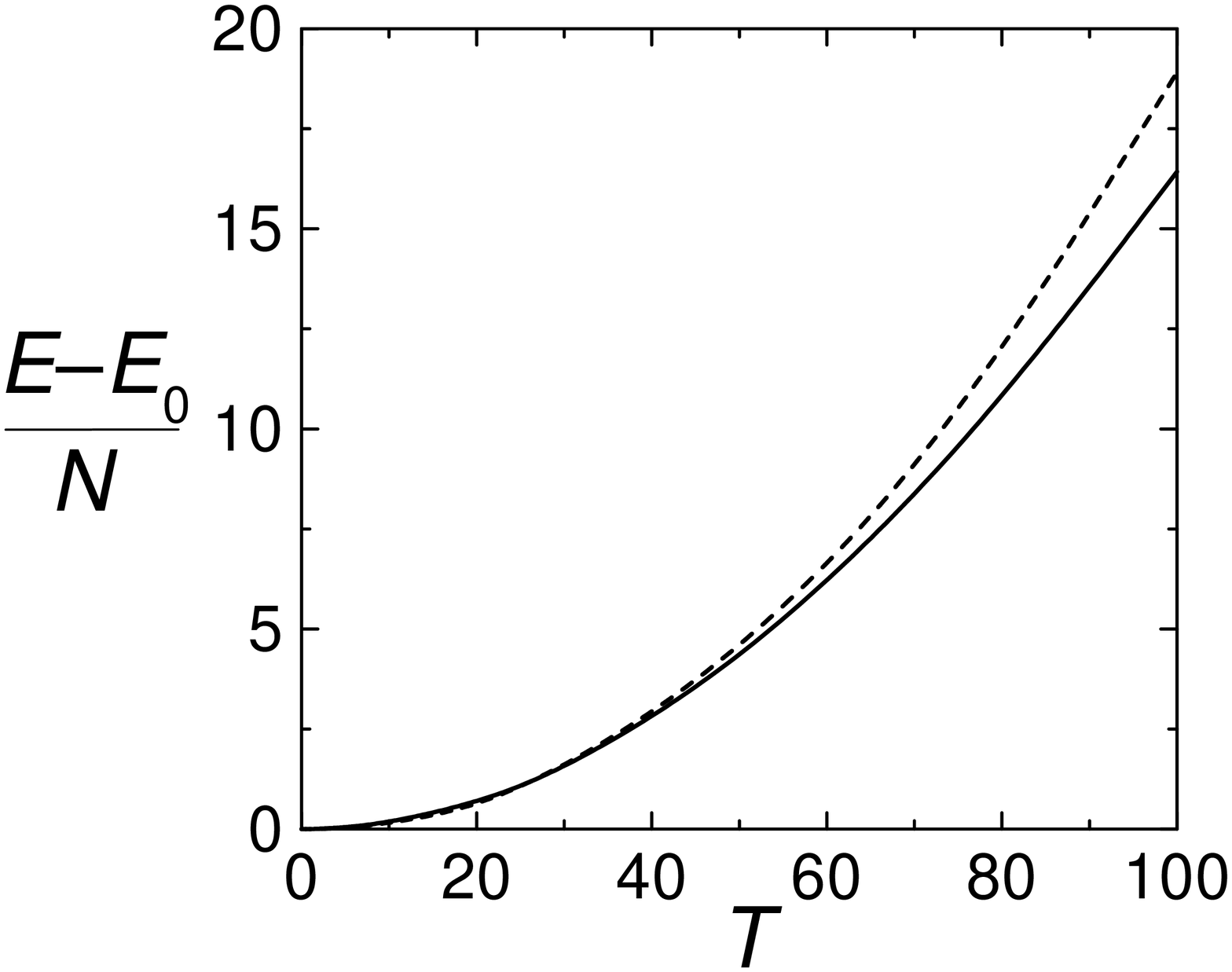}}
\caption{(left) --- Condensate fraction $N_0/N$ as a function of temperature $T$;
(right) --- energy as a function of temperature.
Dashed line is the interacting system, solid line is the ideal Bose-gas.
}
\label{fig1}
\end{figure}

In Fig.~\ref{fig1} the dependencies of the condensate fraction $N_0/N$ and energy $(E-E_0)/N$
on temperature are given for the 1D gas.
For simplicity, the following values of parameters are used in the calculations:
$
\hbar=\omega = 1,\ m=2\pi^2,\ N=1000,\ g=0.0002.
$


Since $c_{nn}>0$, the interaction shifts the energy levels up, which effectively
corresponds to lighter particles. Thus, the Bose-condensation temperature increases.
Note, however, that the presented approach is not valid for higher
temperatures as more items must be included in the Hamiltonian (\ref{H4}).


\bigskip
To summarize, the Bogoliubov's idea of an approximate second quantization approach is generalized for a weakly-interacting
Bose-gas confined in a harmonic trap. As a result, the equations permitting to diagonalize the Hamiltonian and to calculate
the elementary excitation spectrum are obtained. A perturbative solution is applied and the condensate fraction and
energy dependencies on temperature are calculated for a model system to demonstrate the validity of the approach.


{\bf Acknowledgements}.\ \
I am grateful to my colleagues from the Dept.\ for Theoretical
Physics,
Natl.\ University of Lviv, for valuable comments and discussions,
in particular to Prof.~Ivan Vakarchuk, Prof.~Volodymyr Tkachuk, Taras Fityo,
Yuri Krynytskyi.
I wish also to express my gratitude to Prof. Beatrice Meini, University of
Pisa, and to Dr. Hiroaki Mukaidani, Hiro\-shi\-ma University, for their helpful
comments on the matrix equations treatment.



\begin{thebibliography}{00}
\bibitem{And95}M. H. Anderson et al., {\it Science} {\bf 269}, 198 (1995).

\bibitem{Dav95}K. B. Davis et al., {\it Phys. Rev. Lett.} {\bf 75}, 3969  (1995).

\bibitem{Pet01}C. J. Pethick and H. Smith, {\it Bose--Einstein Condensation in Dilute
Gases} (Cambridge, Cambridge University Press), (2001).

\bibitem{Bog47}N. N. Bogoliubov, {\it J. Phys. USSR} {\bf 11}, 23  (1947).

\bibitem{Bas64}W. H. Bassichis and L. L. Foldy, {\it Phys. Rev.} {\bf 133}, A935 (1964).

\bibitem{Oku03}M. Okumura and Y. Yamanaka, {\it Phys. Rev. A}  {\bf 68}, 013609 (2003).

\bibitem{Dal99}F. Dalfovo et al., {\it Rev. Mod. Phys.}  {\bf 71}, 463 (1999);
%
J. O. Andersen, {\it Rev. Mod. Phys.}  {\bf 76}, 599 (2004);
%
R. Ozeri et al., {\it Rev. Mod. Phys.}  {\bf 77}, 187 (2005).


\bibitem{HFu03}H. Fu, Y. Wang and B. Gao, {\it Phys. Rev. A}  {\bf 67}, 053612 (2003).

\bibitem{Kol00}E. B. Kolomeisky et al., {\it Phys. Rev. Lett.} {\bf 85}, 1146  (2000).

\bibitem{Kra96}W. Krauth, {\it Phys. Rev. Lett.} {\bf 77}, 3695  (1996).

\bibitem{Pea98}S. Pearson, T. Pang and C. Chen, {\it Phys. Rev. A} {\bf 58}, 1485 (1998).

\bibitem{Dru97}N. J. van Druten and W. Ketterle, {\it Phys. Rev. Lett.} {\bf 79}, 549  (1997).

\bibitem{Khe05}K. V. Kheruntsyan et al., {\it Phys. Rev. A} {\bf 71}, 053615  (2005).

\bibitem{Bat05}M. T. Batchelor et al., {\it J. Phys. A} {\bf 38} 7787  (2005).

\bibitem{Bag91}V. Bagnato and D. Kleppner, {\it Phys. Rev. A} {\bf 44}, 7439  (1991).


\bibitem{RicattiEqs}
               N. Bessis and G. Bessis, {\it J. Math. Phys.} {\bf 38} 5483 (1997);
               B. Meini, {\it Lin. Alg. Appl.} {\bf 413} 440 (2006);
               H. Mukaidani and T. Shimomura, {\it J. Math. Analysis Appl.} {\bf 267}, 209 (2002).

\end{thebibliography}
\end{document}